# SURVEY OF SEARCH AND REPLICATION SCHEMES IN UNSTRUCTURED P2P NETWORKS

**Sabu M. Thampi**

Department of Computer Science and Engineering, Rajagiri School of Engineering and Technology.

Rajagiri Valley, Kakkanad, Kochi- 682039, Kerala (India)

Tel: +91-484-2427835    E-mail: sabum@rajagiritech.ac.in

**Chandra Sekaran. K**

Department of Computer Engineering, National Institute of Technology Karnataka.

Surathkal-575025, Karnataka (India)

Tel: +91-824-2474000    E-mail: kch@nitk.ac.in

**Abstract**

P2P computing lifts taxing issues in various areas of computer science. The largely used decentralized unstructured P2P systems are ad hoc in nature and present a number of research challenges. In this paper, we provide a comprehensive theoretical survey of various state-of-the-art search and replication schemes in unstructured P2P networks for file-sharing applications. The classifications of search and replication techniques and their advantages and disadvantages are briefly explained. Finally, the various issues on searching and replication for unstructured P2P networks are discussed.

# 1. Introduction

Computing has passed through many transformations since the birth of the first computing machines. A centralized solution has one component that is shared by users all the time. All resources are accessible, but there is a single point of control as well as a single point of failure. A distributed system is a group of autonomous computers connected to a computer network, which appears to the clients of the system as a single computer. Distributed system software allows computers to manage their activities and to share the resources of the system, so that clients recognize the system as a single, integrated computing facility. Opportunity to attach components improves availability, reliability, fault tolerance, and performance. In such systems, the methods for minimizing communication and computation cost are significant. The widely used client–server model is an example of a distributed system. In this model, the servers are optimised to offer services to several clients. The clients always communicate with the servers and they do not share any services. If the servers fail, the whole services that are offered from the servers to the clients are terminated.

The World Wide Web (WWW) can be viewed as a massive distributed system consisting of millions of clients and servers for accessing associated documents. Servers preserve collections of objects, whereas clients provide users a user-friendly interface for presenting and accessing these objects. The inadequacy of the client–server model is evident in WWW. Being resources are concentrated on one or a small number of nodes and to provide 24/7 access with satisfactory response times, complicated load-balancing and fault-tolerance algorithms have to be employed. The same holds right for network bandwidth, which adds to this tailback situation. These two key problems inspired researchers to come up with schemes for allocating processing load and network bandwidth among all nodes participating in a distributed information system [19].

P2P networks are a recent addition to the already large number of distributed system models. P2P systems—an alternative to conventional client–server systems— mostly support applications that offer file sharing and content exchange like music, movies, etc. The concept has also been effectively employed for distributing computing and Internet-based telephony. A major benefit of P2P file sharing is that these systems are fully scalable—each additional user brings extra capacity to the system. A node (peer) can act both as a client and a server. The participating nodes mark at least part of their resources as 'shared', allowing other contributing peers to access these resources. Thus, if node A publishes something and node B downloads it, then when node C asks for the same information, it can access it from either node A or node B. As a result, as new users access a particular file, the system's capability to provide that file increases [1].

There are mainly three different architectures for P2P systems: centralized, decentralized structured and decentralized unstructured (Fig 1). In the centralized model, such as Napster [2], central index servers are used to maintain a directory of shared files stored on peers with the intention that a peer can search for the location of a desired content from an index server. On the other hand, this design makes a single point failure and its centralized nature of the service creates systems susceptible to denial of service attacks. Decentralized P2P systems have the advantages of eliminating dependence on central servers and providing freedom for participating users to swap information and services directly between each other. In decentralized structured models, such as Chord [3], Pastry [4], and CAN [5], the shared data placement and topology characteristics of the network are strongly controlled on the basis of distributed hash functions. In decentralized unstructured P2P systems, such as Gnutella [6] and KaZaA [7], there is neither a centralized index nor any strict control over the network topology or file placement. Nodes joining the network, following some loose rules, form the network. The resulting topology has certain properties, though the placement of objects is not based on any knowledge of the topology [8]. The decentralization makes available the opportunity to utilise unused bandwidth, storage and processing power at the periphery of the network. It diminishes the cost of system ownership and maintenance and perks up the scalability.

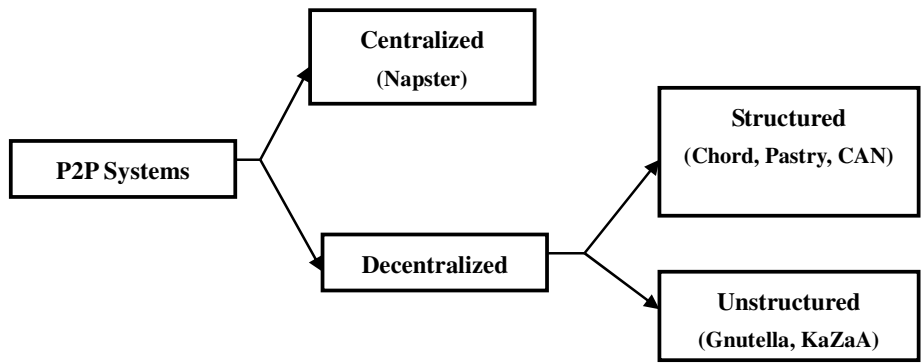

**Figure 1. Different Architectures for P2P Systems**

Gnutella is mainly used for file sharing. Like most P2P systems, Gnutella builds, at the application level, a virtual overlay network by means of its own routing technique [9]. As a result, it functions as a distributed file storage system, permitting its clients to spell out directories on their machines, which they want to share with other peers. Since it is a purely decentralized architecture, there is no central coordination of the activities in the network and users connect to each other directly through a software application which functions both as a client and a server and is therefore referred to as a servent [10]. Fig. 2 presents visualizations of the Gnutella network topology [11]. A servent opens one or many connections with nodes that are already in the network to become a member of the network. As nodes often join and leave the network, connections are unreliable. To deal with this situation, after joining the network, a peer occasionally pings its neighbours to find out other participating nodes. Using this information, a disconnected node can always reconnect to the network. Nodes decide where to connect in the network on the basis of local information only and thus form an active, self-organizing network of autonomous entities [11].

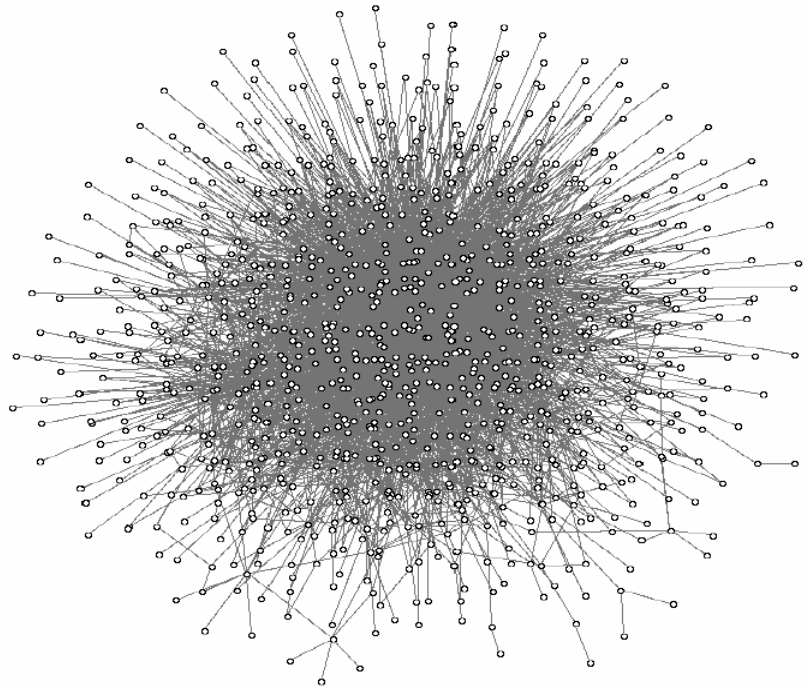

**Figure 2. A Snapshot of Gnutella Network on Dec. 28, 2000** (from [15])

KaZaA [7] is a P2P file-sharing application, which employs the idea of 'superpeers' (Fig. 3). The nodes form a structured overlay of superpeers, which are nodes with more bandwidth, disk space and processing power. The ordinary peers that do not have a lot of storage, processing or communication resources transmit the metadata of the files they are sharing to the superpeers. All the queries are forwarded to the superpeers.

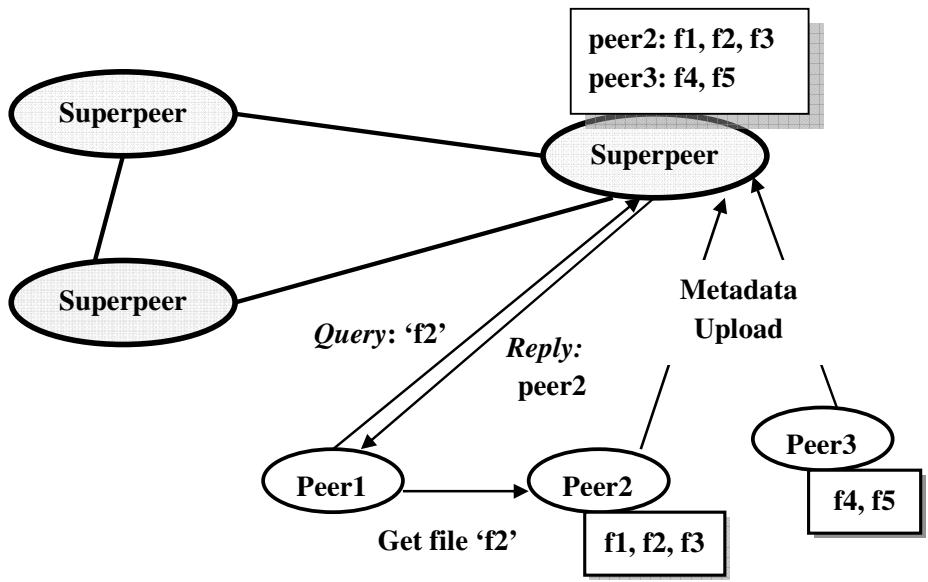

**Figure 3. Resource Discovery in Superpeer Architecture**

The scale of P2P systems and the absence of centralized control create complex performance issues such as bandwidth utilization, network traffic control, and security. Many techniques developed in the past for distributed systems of a few number of servers may no longer apply; innovative methods are required to meet the challenges in P2P computing [14]. Given the importance of various problems, several researchers have attempted to tackle the issues and developed numerous solutions. Some of the techniques are very simple, while others employ complex procedures, all leading to different levels of efficiency [15]. Most of the research in the field of P2P file-sharing systems focuses on developing efficient search, replication and security techniques. In addition, there exist some important research areas for data-sharing systems such as resource management issues that include fairness and administrative ease [14].

As the nodes in a P2P network bank on resources received from each other, efficient searching and availability guarantee are vital in P2P networks. A search process includes aspects such as the query-forwarding method, the set of nodes that receive query-related messages, the form of these messages, local processing, stored indices and their maintenance. Designing a good search mechanism is difficult in P2P systems for several reasons, including scale of the system and unreliability of individual peers [49]. In a P2P network, the quality of query results is measured by the number of results and efficiency in object discovery measures search accuracy and the number of discovered objects per request. Bandwidth consumption is an essential attribute, as it gives users a much broader choice for object retrieval. The routing efficiency is normally measured by the number of overlay hops per query. In some systems, it is also assessed using the number of messages per query. The search effectiveness decays exponentially as the search time augments because the number of query messages increases linearly with the volume of visited peers. It is also vital that any search algorithm adapts to dynamic environments, since in most P2P networks users usually enter and leave the system frequently. Different searching techniques make dissimilar trade-offs between these desired characteristics [8, 17, 18, 19].

During searching, several query packets pass through the network searching for the target objects. The heterogeneity of these query packets creates a local traffic disparity and congestion. The downloading of large objects in response to requests also causes congestion in nodes. One proficient method for forestalling this load concentration is replication of the target objects into various sites. Replication increases object availability and fault tolerance. Single node failures, like crashes of nodes, can be tolerated as faults within the system as a whole facilitated with the help of the redundancy introduced by replicas. If a host of a replica fails, requestors may access another host with a replica. Data replicated at more than one site facilitate to minimise the number of hops before the data are found. To maintain consistency of replicated data, there are some P2P-specific challenges to overcome, which include high rates of peer failures, node selection logic for hosting the replica and lack of global knowledge on shared data [20].

Placing a large quantity of data on every node or flooding a query to every peer significantly diminishes the performance and effectiveness of the system. Therefore, the placement of data into various nodes (replication) and routing requests through nodes (searching) in the network should be done efficiently. Several search and replication techniques are proposed in the literature. This paper theoretically reviews some of those techniques for unstructured P2P networks. The advantages and disadvantages of all the search and replication techniques considered in this paper are summarised in separate tables.

The paper is organised into six sections. Section 2 and 3 respectively review the existing search and replication techniques for unstructured P2P networks. Section 4 presents an overview of previous work. Section 5 discusses various issues on search and replication processes. Finally, section 6 concludes the paper.

## 2. Search Techniques

The objective of a search mechanism is to successfully locate resources while incurring low overhead and delay [21]. With the aim of fulfilling the goal, a number of search algorithms have been proposed for effective resource discovery in decentralized P2P networks. Search methods for decentralized unstructured P2P networks are categorized in different ways. One classification is based on query forwarding: deterministic and probabilistic [17]. In a deterministic approach, the query forwarding is deterministic (e.g. local indices). Prior information on query path is used for routing. In the probabilistic forwarding, the query is routed either probabilistically or randomly. Random walk methods route queries through randomly selected nodes.

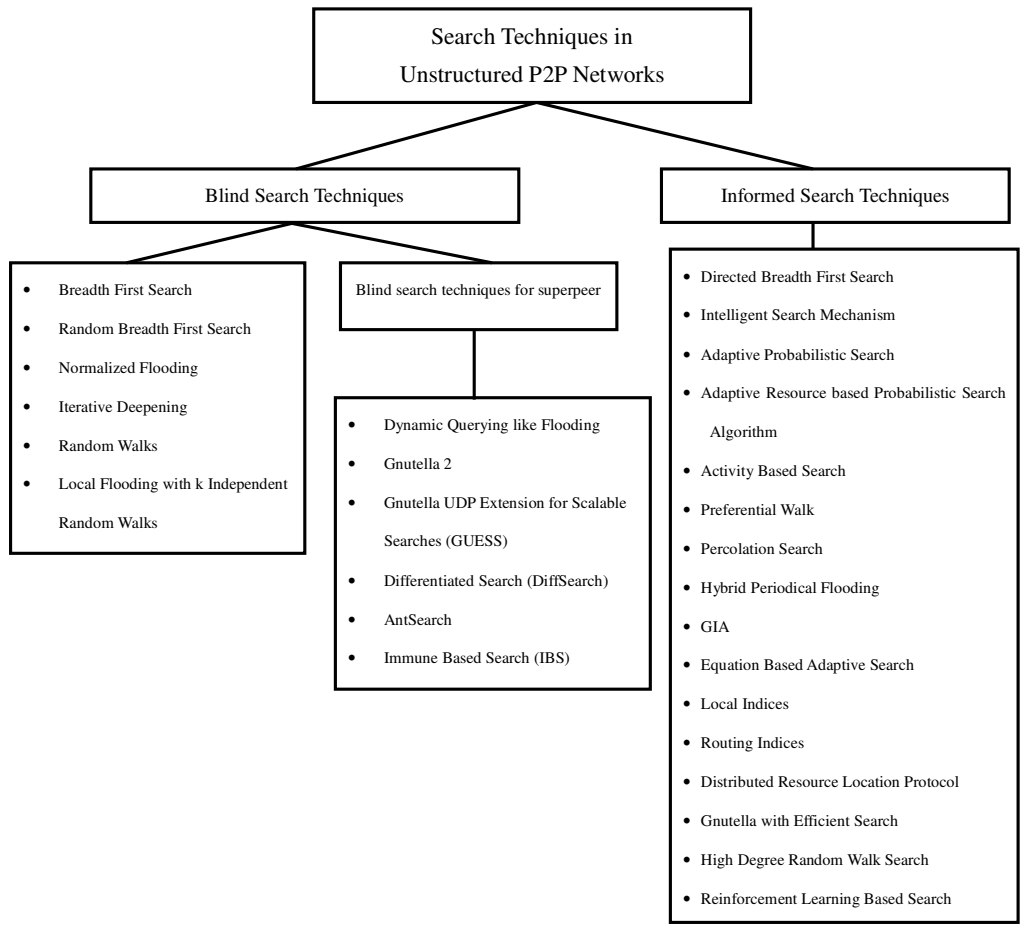

**Figure 4. Search Techniques in Unstructured P2P networks**

The second arrangement is blind search and informed search [17, 19, 22]. This classification (Fig. 4) stands on exploitation of location information of peers or objects. In a blind search, nodes do not keep information about object location. In an informed search, nodes gather some metadata that assist the search operation. There are a few blind search techniques associated with super node–based architecture for minimising the expected search cost. These schemes are usually called controlled flooding techniques. There are seven blind search techniques, six controlled flooding techniques and sixteen informed search techniques are discussed in this paper.

*2.1 Blind Search Techniques*

Blind search schemes employ flooding techniques to relay queries to peers in the network. Peers keep no information about the P2P network or the probable locations of objects for routing queries. The attributes such as scalability, load-balancing, success rate etc are employed to evaluate the performance of a search technique. The Table 1 summarizes the query forwarding mechanism and the advantages and disadvantages of each of the blind search technique.

*Breadth First Search (BFS):* BFS is widely used in file-sharing P2P systems such as Gnutella [6]. BFS starts at the query node by examining every neighbour if it is the target node. If this fails, each of these neighbours tries their neighbours and this goes on until the target peer is found. While the searching in is very simple, each query guzzles too much network and processing resources for the reason that queries spread along all links. Thus, low-bandwidth nodes simply turn into a bottleneck [23, 24]. Hence, a time-to-live (TTL) scheme is employed to limit flooding of queries. Each query is propagated with a TTL and query is terminated when either result is found or TTL is exhausted. Though this flooding method still generates enormous amount of overhead by contacting several peers, it guarantees high success rate.

*Random Breadth First Search (RBFS):* RBFS [24, 42] is a blind search technique that has been proposed as an alternative to traditional flooding scheme (BFS). The idea behind RBFS is to improve on the flooding scheme to decrease communication overhead during search. It utilises only local connectivity information of the network during search. It randomly selects a portion of neighbours that are visited for each node. Search then continues from the query node examining only n neighbouring peers and proceeding forward. The portion of peers that are chosen is a parameter to the search mechanism. RBFS improves upon BFS by reducing the number of messages passed during search. RBFS is probabilistic and still visits more peers.

*Normalized Flooding:* Normalized flooding [26] is the same as flooding (BFS), except that every node sends the message only to a subset of its neighbours. Let 'δ' be the minimum degree of a node in the network. If a node has more than δ neighbours, then it sends the message only to 'δ' nodes in its neighbourhood that are selected uniformly at random. The low degree nodes hosting good content are left out during searching if 'δ' is not properly selected. Other features of nodes such as number of objects, storage and bandwidth are not considered in this technique for routing.

*Iterative Deepening (expanding ring):* Iterative deepening [17, 22, 27] - an idea borrowed from AI-is depth-first search to a fixed depth in the tree being searched. The querying node occasionally sends out a series of BFS searches with increasing depth limits $D_1 < D_2 < D_i$. The query is finished when the query result is satisfied or when the maximum depth limit, D, has been reached. In the latter case, the query result may not be satisfied. All nodes use the same sequence of depth limits called policy P and the same time period between two consecutive BFS searches. Iterative deepening is customized to applications where the initial number of data items returned by a query is significant. The main shortcomings of iterative deepening are creation of large amount of duplicate messages and slow query processing.

*Random Walks:* Random walk, also known as Markov chain and 'drunkard's walk', is a path constructed by taking successive steps in random directions. The Markov property means the system is memory-less, i.e. previous states are irrelevant for predicting the subsequent states [15]. The idea of random search is that instead of forwarding a request to all the neighbours, the recipient sends it to a random neighbour. This search mechanism does not generate as much message traffic as the BFS algorithms since there is only one message being routed in the system. The trade-off is that the search response time is significantly longer. There are variants of the random walk method, which intends to diminish the search time. One form involves the use of k independent random walks [8]. The search is successful once the object is found by any one of the individual random walks. This method of searching objects is called k-walkers algorithm [26]. A walker terminates with a success or failure. Two different methods are used for terminating the walkers: TTL-based termination and checking-based termination. After traveling a certain number of hops, the random walk is terminated in TTL-based method. In the case of 'checking', the walker periodically contacts the query node before moving to next node. While k-random walk approach manages to reduce messages appreciably, it shows low performance because of its random nature and incapability to adjust to different query loads.

Another variant of random walk is a two-level random walk [28]. It uses a two-level random walk policy. The querying node selects $k_1$ random walks with $TTL_1 = l_1$. When the $TTL_1$ finishes at a particular node, each thread will then generate k2 threads which will perform $k_2$ random walks from that node with $TTL_2 = l_2$. Given the equal number of walkers, this scheme generates less duplicate messages but has longer searching delays than the k-walker random walk.

*Local Flooding with k Independent Random Walks:* Local flooding with k independent random walks [26] searching method is a conciliation between flooding and random walks. The idea is to first perform a local flooding starting from the query source; the flooding goes on until precisely k new outer nodes have been discovered, for some predefined value of k. In the case where one of these nodes hosts the object, the search is successful and the query source is informed. If not, each of the k nodes begins an independent random walk. If the file is located close to the origin, the local flooding would be sufficient to locate it fast, and with few messages exchanged. If the file is located away from the network, it is expected that it will be located by one of the random walks. Since the flooding occurs only locally, the message complexity is small. If the search continues by independent random walk after local flooding with large TTL values, message production will be high and the performance will be degraded.

**Table 1. Advantages and Disadvantages of Blind Search Techniques**

| Search Technique | Query forwarding | Advantages | Disadvantages |
|---|---|---|---|
| Breadth First Search | Flooding of queries with certain TTL | i. Finds out rare objects.<br>ii. Performs very well for small TTL values for high replication. | i. Generates large amount of unnecessary traffic, and thus wasting bandwidth and processing resource.<br>ii. Does not scale well. |
| Random Breadth First Search | Random selection of target nodes for query forwarding. | Message production is less as compared to BFS, but still contacts a large number of peers. | Success rate deteriorates since target nodes are selected randomly and high performing nodes may be omitted. |
| Normalized Flooding | Target nodes are selected based on the minimum degree of nodes in the network. | Nodes send the messages to at most 'm' neighbours. Hence, message overhead is reduced. | i. Random selection of nodes.<br>ii. Low degree nodes are left out from receiving the queries. |
| Iterative Deepening | The search process involves initiating successive multiple breadth-first searches with increasing depth limits. | i. Attains good success rate when the search termination condition links to a user-defined number of hits.<br>ii. Improves scalability.<br>iii. Performs well when popular, well-replicated items are to be discovered. | i. Imposes substantial overhead, by propagating through the substantially larger portion of the same nodes each new cycle.<br>ii. Creates large amount of duplicate messages<br>iii. Slow query processing. |
| Random Walks | The number of nodes to which query is forwarded by each node is limited to a number 'k'. The 'k' nodes are selected randomly. | i. Small message complexity (algorithm scales well with the size of the network).<br>ii. Attains local load-balancing as all the nodes in the network are treated as equal. | i. Variable performance, hence success rate and number of hits vary due to random choice of neighbours for routing queries.<br>ii. The peers available for a long time could be chosen more often.<br>iii. No explicit technique is used to guide a search query.<br>iv. Queries for popular and unpopular objects are propagated in the same way. |
| Local Flooding with k Independent Random Walks | Conciliation between flooding and random walks. | i. Combines the advantages of flooding and random walk.<br>ii. The message complexity is small, if flooding occurs locally. | Message overhead is high if walkers travel more hops. |

*2.2 Blind Search Techniques for Superpeer Structure*

A few blind search techniques for superpeer structure based on controlled flooding are introduced recently. They are Dynamic Querying like Flooding, Immune Based Search, Gnutella 2 protocol, Differentiated Search, and GUESS.

*Dynamic Querying like Flooding:* Dynamic querying (DQ) is a new flooding technique for superpeer architecture. The scheme could estimate a proper time-to-live (TTL) value for a query flooding by estimating the popularity of the searched files, and retrieve sufficient results under controlled flooding range for reducing network traffic [30]. In ultrapeer based unstructured P2P networks, this type of controlled flooding endeavors in locating an object at the minimum message cost. Nevertheless, it *severely increases the latency perceived* by the peers. The enhanced algorithm for DQ - *Dynamic Querying like Flooding* [31] technique contains all features of dynamic querying, which mainly includes probe queries and TTL computation. The source peer starts the search (for N results) by first sending a query towards a small number of neighbours with a fixed TTL; this is just like in the original dynamic querying. The network replies $n \geq 0$ results that can be used to infer the popularity of the item. After getting an estimate of its popularity, it then calculates the TTL for the next neighbour. A query packet with this TTL value is spread via the next neighbour. The new number of results is obtained and iterated to calculate the TTL for another neighbour. This process continues until the desired number of results is obtained or all neighbours are used. This enhanced greedy and conservative flooding algorithm reduces the latency by more than four times, and loosens the constraints on the network connectivity. Low degree peers often find the right number of results at low search cost. The search cost of less popular items is high.

*Immune Based Search (IBS):* A large amount of redundant messages is generated in DQ-like search techniques. The IBS [32] algorithm is a controlled flooding algorithm to search results for a query with a specified required number of results. It has been inspired by the concept of the biology immune system where B cells undergo mutation and opportunistic proliferation to generate antibodies, which track the antigens. IBS has several salient features by utilizing the immune method. The search process starts at the query source Q. The query source carries out a few initializations subsequent to probe phase. First, it assigns TTL with a value, which is attained from the probe phase. Second, it provides Q a unique identifier, denoted by QID, which is made of a unique peer-id and a query counter managed by the query source. Peers use QID to differentiate between fresh queries and those received before. Third, the core of Q is the search profile PS of the user peer. After initialization, the query originator generates the sequence of the other four sub-phases: query forward, local query execution, merge-and-backward, and data retrieval. It is implemented independently by each individual peer participating in the network and is totally decentralized in nature. ImmuneSearch avoids query message flooding; instead it uses an immune systems inspired concept of affinity governed proliferation and mutation for message movement. In addition, a protocol is formulated to change the neighbourhoods of the peers based upon their proximity with the queried item. The topology evolution coupled with proliferation and mutation help the P2P network to develop 'memory', as a result of which the search efficiency of the network improves as more and more individual peers perform search. The search latency in IBS is less than DQ algorithm. IBS algorithm can estimate the reasonable TTL base on probe phase and efficiently utilize the history information to guide the query forwarding, consequently query latency is decreased. However, the higher churn rate of peers causes, reduction in the performance of IBS.

Another technique which employs the swarm intelligence Bees Algorithm is called P2PBA (Peer to Peer file sharing - Bees Algorithm). This scheme is based on the lines of food search behaviour of Honey Bees [73]. The scheme optimizes the search process by selectively going to more promising honey sources and scan through a sizeable area.

*AntSearch:* A recent study on Gnutella file sharing system shows that as many as 70% of its users do not share any files at all. This means that these users use the system for free. This behaviour of an individual user who uses the system resources without contributing anything to the system is the free riding problem. Such users are referred to as *free riders* [33]. The free-riders problem causes a large amount of redundant messages during searching. AntSearch [30] manages the problem of freeriding. In AntSearch, each peer maintains its success rate of previous queries and records a list of pheromone values of its immediate neighbours. Based on the pheromone values, a query is only flooded to those peers which are not likely to be the freeriders. The search process comprises a probe phase and a flooding phase. In the probe phase, the requester peer obtains the statistics information about the searched files after flooding queries to a few neighbours with a small TTL. The statistics information is stored into a data structure called the "probe table". In the flooding phase the requester peer calculates how many peers should be further contacted, and chooses a suitable TTL for a neighbour. The requester peer then propagates the query packet towards a neighbour, and all the following peers only forwards the query to the k% of neighbours with higher pheromone values. This iterative process stops when the desired number of results is returned, or all neighbours have been visited. The technique significantly reduces the redundant messages during a query flooding and on the other hand, the search latency is same as DQ like search algorithm. AntSearch sometimes retrieves a larger number of results than that in DQ. This overshooting problem is caused because the physical number of searched peers is larger than the estimated number of peers.

*Gnutella 2:* Gnutella2 [34, 35, 36] divides nodes into two groups: leaves and hubs. Leaves keep one or two links to hubs, whilst hubs allow hundreds of leaves, and numerous connections to other hubs. When a search is initiated, the node contacts the hubs in the list, noting which have been searched, until the list is exhausted, or a predefined search boundary has been reached. This permits a user to locate a popular file without loading the network. Hubs index what files a leaf has using a Query Routing Table. Gnutella2 also has a metadata system for more complete labeling, rating, and quality information to be given in the search results. Gnutella2 utilises compression in its network connections to reduce the bandwidth used by the network. Gnutella 2 is more efficient, as continuing a search does not augment the network traffic exponentially, queries are not routed through as many nodes, and it increases the granularity of a search, allowing a client to stop once a pre-defined threshold of results has been obtained more effectively than in Gnutella. However, the scheme increases the complexity of the network and the network maintenance required due to extra indices to be maintained in super peers.

*Gnutella UDP Extension for Scalable Searches (GUESS):* In GUESS [37, 38, 39, 40], a search is carried out by iteratively contacting various super nodes and having them inquire all their leaves, until sufficient objects are found. The important characteristic of GUESS is that query messages are not broadcasted via flooding-based routing. As an alternative, a peer merely iterates through the entries in its link cache, and performs a search to the target peer. Peers running the GUESS protocol maintain two caches: a link cache, and a query cache. A peer should search only as many other peers as needed to get sufficient number of results. As the size of the cache increases, there are more peers to probe. In order to reduce message forwarding and augment the number of discovered objects, the number of leaves per superpeer must be kept high. The process of selecting peers for probing is not conferred in GUESS protocol. Although larger cache sizes result in a larger number of probes, they do not translate to more satisfied queries. Since there are only limited number of peers that tend to share a large number of objects, many queries will go unsatisfied.

*Differentiated Search (DiffSearch):* DiffSearch [41] algorithm, which is based on ultrapeer overlay, improves the search efficiency of unstructured P2P networks by giving higher querying priority to peers with high querying reply capabilities. In the DiffSearch algorithm, a query consists of two round searches. In the first round search, the query is only sent to the ultrapeer overlay. If the first round search fails in the ultrapeer overlay, the second round search will be evoked to query the entire network. The prerequisite of the DiffSearch algorithm is that the ultrapeer overlay consisting of content-rich peers is well formed in a P2P network. To solve the load- balancing problem, indices of leaf nodes to ultrapeers should be uploaded to the ultrapeers. The advantage of differentiated search is that the search traffic is significantly reduced due to the shrunken search space. However, the second round of search causes overhead in the network due to flood of messages. Moreover, the index uploading adds a small cost to the overhead.

## 2.3 Informed Search Methods

In informed search mechanisms, peers maintain some kind of routing information to forward queries to suitable peers. This information are based on several parameters such as popularity of objects, success rate etc. Informed search approaches offer smaller response time than blind search approaches but at the cost of increased overhead of maintaining various indexes. The Table 2 lists the query forwarding mechanism and the various advantages and disadvantages of different informed search techniques.

*Directed Breadth First Search (DBFS):* In DBFS scheme [27], a query source sends query messages to just a subset of its neighbours. The neighbours that receive the query then continue forwarding the message to all neighbours as with BFS. To facilitate intelligent selection of neighbours, a node maintains information on its neighbours. These data include the number of results that were received through the neighbour for previous queries or the latency of the connection with that neighbour. Since queries are transmitted through a small subset of neighbours, the number of nodes that receive the query are significantly reduced.

*Intelligent Search Mechanism(ISM):* ISM [23, 25], which is also called intelligent-BFS, is an informed search algorithm framed with the objective of facilitating the querying peer to locate the most appropriate answers to its query effectively. Each peer in the network maintains a profile of its neighbours, and utilises the profile to locate the target peers which are likely to answer for a query. It then forwards the query to those peers only. The profiles are aggregated and collected in real-time. The performance of the ISM improves over time as the peers learn more information about their peers and therefore becomes better than the RBFS. ISM works well in environments which exhibit strong degrees of query locality and where peers hold some specialized knowledge [23]. ISM focuses more on object discovery than message reduction. The method fails to consider adaptation to peer departures and it does not make use of negative feedback. The accuracy of search results depends very much on the assumption that nodes specialize in certain documents [22].

*Adaptive Probabilistic Search (APS):* APS [19, 22, 43] utilises quantitative data in the form of probabilistic information for the purpose of guiding search operations. The main difference with random walkers is that in APS a node makes use of response from earlier searches to probabilistically direct future walkers, instead of forwarding the walker at random. In APS, each node maintains a table for the forwarding probability to each neighbour for each resource. The value of each entry in the table mirrors the relative probability of this node's neighbour to be selected as the next hop in a future request for the specific object. APS employs k random walkers to search for the required resource and each intermediate node forwards the query to one of its neighbours with a probability given by its table index. Index values are updated after each query using the feedback from walkers and the update procedure takes the reverse path back to the requester. If a walker succeeds (fails), the relative probabilities of the nodes on the walker's path are increased (decreased). APS considers duplicate messages as failure states. The search algorithm shows improved performance over the random walker model. APS achieves high success rates, increased number of discovered objects, low bandwidth consumption and adaptation to changing topologies. The main disadvantage of APS is the probabilistic selection of nodes for query forwarding without employing various peer parameters such as bandwidth, storage availability etc. for choosing the target peers.

*Adaptive Resource-based Probabilistic Search Algorithm (ARPS):* ARPS [44] introduces weighted probabilistic forwarding for query messages on the basis of node degree distribution and popularity of the resource being searched. The search scheme employs a method to estimate the popularity and adjust the forwarding probability consequently. Every node maintains a local index on the popularity estimate for every resource it has demanded or forwarded queries. On the basis of the estimation, a proper forwarding probability is selected to spread the query messages. Peers forward the query messages to its neighbours with small weighted probability if the resource being searched is with high popularity, and vice versa. The weighted probability is smaller for high degree nodes in contrast to low degree ones. During a search process, a node first locates the resource in its local index. If there is, no match the peer floods the queries to its neighbours with a forwarding probability of 1; or else, a reasonable probability is chosen to forward the message. Authors claim that ARPS scheme guarantees popularity-invariant search success rate. The memory requirement is less than that of APS. The performance of ARPS depends on the decay rate to some extent. The flooding spawns abundance of messages in the network.

*Activity Based Search (ABS):* ABS [37] is a scalable search algorithm for dynamically find out the number of nodes to forward a query. The decision is based on the degree to which each neighbour has contributed to a successful search, called the *activity level*. Every node dispatches a query to $d_K$ neighbours with the highest activity levels at once. A node merely considers the previous success ratio to decide $d_K$ and the nodes to send a query to. The algorithm creates a spanning graph consisting of high traffic links. Once a stable spanning graph is formed, the number of hops required for a search is approximately bound by the diameter of the spanning graph. The algorithm exhibits notable performance than both *k*-random walk and *1*-random walk. However, the forwarding of queries to peers with highest activity levels increases load on them and thus performance is diminished.

*Preferential Walk:* Preferential walk (P-walk) [45] is a trust-based probabilistic search algorithm which utilises a trust evaluation method to rate neighbours according to the feedback from previous searches. Every peer ranks its neighbours according to the search experience. The highly ranked neighbours have higher probabilities to be queried. Neighbouring peers assign each other trust ranks. During routing, peers preferentially forward queries to the highly ranked neighbours, thus messages are routed to most likely peers, which hold the desired resources. Appropriate measures for handling duplicate messages, free riding and partial coverage are not addressed in P-walk.

*Percolation Search:* Percolation search [46] is a decentralized algorithm that is able to locate any item in a P2P network with minimal search traffic. The search scheme has the ability to answer complex queries using scalable resources. The key steps in the search algorithm are (i) *content caching*: an initial one-time-only replication of a node's content list or directory in the nodes visited via a short random walk, (ii) *query implantation*: when a node wants to make a query, it first executes a short random walk and implants its query request on the nodes visited and (iii) *bond percolation*: when the search begins, each node having the query implantation starts a broadcast search; conversely it only sends a query to a neighbour with probability q. In this search scheme, the high capability nodes evolve into a sub-graph with a heavy tail degree distribution and hence will carry the majority of the search load. The peers available for a long time could be chosen more often and thus other nodes may be left out from routing queries.

*Hybrid Periodical Flooding:* Blind flooding (BFS) among peers causes huge volume of superfluous traffic and considerably minimizes the query coverage range. This is called the partial coverage problem. Hybrid periodical flooding (HPF) [29] minimizes the partial coverage problem and reduces traffic than BFS. HPF provides the suppleness to adaptively fine-tune various parameters to meet diverse performance requirements. The most important feature of HPF is that the number of relay neighbours can be changed periodically on the basis of a periodical function and they are selected on the basis of multiple metrics in a hybrid way. There are several metrics that can be used to choose relay neighbours, for instance, communication cost, bandwidth, number of returned results from the neighbour, and average number of hops from the neighbour to peers who responded the previous queries. The principle of iterative deepening is employed to terminate the successful queries. The query processing in HPF is slow since searching employs iterative deepening.

*GIA:* The GIA [47] search protocol uses a biased random walk. To avoid overloading any one node, GIA uses an active flow-control system in which a sender is allowed to direct queries to a neighbour only if that neighbour has notified the sender that it is willing to accept queries from the sender. A GIA node selects the highest capacity neighbour for which it has flow-control tokens and sends the query to that neighbour. If it has no tokens from any neighbours, it queues the query until new tokens arrive. Bookkeeping procedures are used to avoid redundant paths. With bookkeeping, each query is allocated a unique global identifier (GUID) by its source node. A node memorizes the neighbours to which it has previously forwarded queries for a known GUID. If a query with the same GUID arrives back at the node, it is forwarded to another neighbour. This reduces the possibility that a query navigate the same path twice. GIA thus, focuses on how to balance the query load and intends to reduce the query-dropping rate in the flooding. GIA outperforms flooding based search schemes by many orders of magnitude in terms of aggregate query load. All nodes maintain pointers to the content offered by their immediate neighbours to provide one-hop replication. Thus, the indexing of the neighbours' repositories increases the responsibilities of each peer plus communication overhead. Another issue is how fast the algorithm can work for joining peers and at what cost for their neighbourhood.

*Equation Based Adaptive Search (EBAS):* EBAS [48] is an equation-based adaptive search mechanism that uses an estimate of the popularity of a resource to choose the parameters of random walk such that a targeted performance level is achieved by the search. The technique employs a low-overhead method for maintaining an estimate of popularity that utilises feedback obtained from previous searches. EBAS consists of two components: *a parameter selection module* and *a popularity estimator*. The performance of random walk largely depends on the choice of 'k' (number of walkers) and 'T' (TTL limit). The objective of the parameter selection module is to select k and T for discovering a resource with popularity 'p' such that a target performance is guaranteed. Each node maintains a popularity estimate of each resource in a popularity table. The popularity estimate of a resource is updated after feedback from most recent searches for the resource. When a node needs to search for a resource with popularity 'p', it looks up the entry in the parameter selection table corresponding to the interval in which p lies, and initiates random walk with parameters k and T specified in the entry. Therefore, it is ensured that the desired average success rate is achieved while the average overhead and delay are within the specified bounds. This is in contrast to pure random walk whose parameters remain constant, which often leads to low success rate or excessive overhead.

*Local Indices:* In this scheme [49], a node 'n' keeps an index over the data of each node within 'r' hops of itself, where r is a system-wide variable known as the radius of the index. When a node receives a query message, it can then process the query for every node within 'r' hops of itself. In this fashion, the collections of many nodes can be searched by processing the query at few nodes. *A policy* specifies the depths at which the query should be processed. All nodes at depths not listed in the policy simply forward the query to the next depth. To produce and uphold the indices at each node, extra steps must be taken every time a node joins or leaves the network, and each time a user updates his/her local collection. The main difference between a local Indices policy and an iterative deepening policy is that depths in the policy correspond to the depths at which iterations should end, and nodes at all depths process the query. The accuracy and success rate of local indices scheme are high, since each contacted node indexes an entire neighbourhood. Conversely, message production is similar to the flooding scheme. Extra cost for maintaining the index in each node increases the overhead. The scheme also requires a flood with TTL=r on every occasion a node joins/leaves the network or updates its local repository, as a result the overhead turn out to be even bigger for dynamic environments.

*Routing Indices (RIs):* RIs [50] is a hybrid searching technique where each peer builds indices using aggregate information on the contents of the documents of its peers. The RIs kept in a peer gather some information about the data stored at other reachable peers. This information is used to direct the queries towards the peers that hold the required data. The notion of horizons is used to limit the number of peers for which each peer maintains indexing data. If a node cannot respond for a query, it forwards the query to a subset of its neighbours, on the basis of its local RI, rather than by selecting neighbours at random or by flooding. There are three RI schemes: the compound (CRI), the hop-count (HRI), and the exponential (ERI) RIs. Given the index, the 'goodness' of each node for a query is computed. For compound RI, the number of objects that may be found in a path is used as a measure of goodness. The main constraint of the compound RI is that it does not consider the number of 'hops' necessary to find documents. In the hop-count RI, aggregated RIs for each 'hop' up to a maximum number of hops is employed. This number is called the *horizon* of the RI. The hop-count RI is efficient concerning the number of hops. On the other hand, this advantage arrives at a higher storage and communication cost than the compound RI. The hop-count RI performance is pessimistically influenced by the lack of information outside the horizon. The exponential aggregated RI surmounts these inadequacies at the cost of a few potential losses in accuracy. The exponential RI can keep information for all nodes accessible from each neighbour in the RI. Among the RIs, CRI has the best performance, followed by the ERI and HRI. This difference in performance is a function of the number of nodes used to generate the index. In particular, CRI uses all nodes in the network, HRI uses nodes within a predefined a horizon, and ERI uses nodes until the exponentially decayed value of an index entry arrives at a minimum value. This result shows that the more nodes an RI uses to compute the goodness of a path, the better the RI is. The free riding and partial coverage issues are not addressed in none of the schemes of RI.

*Distributed Resource Location Protocol (DRLP):* DRLP [37, 51, 52] behaviour relies on probabilistic parameters that can be adjusted to attain a desired probability value. Every peer has a local directory (LD) that points to locally managed resources such as files. Each peer manages its local resources and its LD and every resource has a unique, location-independent GUID. A node in the network maintains a directory cache (DC) which contains the presumed location of resources managed by other peers. Directory entries travel from LDs to DCs at other nodes to adjust the access patterns. When a peer receives a query message, it searches its LD first and then it's DC. If it finds the object in the LD, it sends a resource found message along the path the search request message traverses until it reaches the query source. This message updates the DC at every peer it visits. If the peer finds the resource in the DC, it sends the search request message to the peer pointed to by that DC. This peer might no longer have the resource, so the search continues from that point forward. If a peer does not find the resource in its LD or DC, it will send the request to each peer in its neighbourhood with a certain probability p, called the *broadcast probability*. The protocol avoids the inefficiency of flood broadcasting methods by using a probabilistic message dissemination method combined with a caching mechanism. By adjusting the probability of routing search request messages, one can vary the probability that the search is successful. The algorithm initially spends many messages to find the locations of an object.

*Gnutella with Efficient Search (GES):* GES [53] utilises a distributed topology adaptation algorithm to arrange semantically relevant nodes into same semantic groups through the concept of node vector. Vector size provides a good trade-off between search performance and bandwidth cost. GES is founded on intuition: if nodes are semantically relevant, it is likely that they are relevant to the same queries. Node vectors are used to compute relevance of nodes. If the relevance score of two nodes' node vectors is high, then the two nodes are semantically relevant. The distributed topology adaptation algorithm is executed frequently at each node to reorganize the overlay such that semantically relevant nodes are organized into same semantic groups through semantic links. Each node also connects to some irrelevant nodes through random links by which GES can find out various semantic groups. The search protocol is a mix of biased walks and flooding. Given a query, GES first uses biased walks through random links to locate a relevant semantic group for the query, and then floods the query through semantic links within the semantic group to retrieve relevant objects. The search continues until sufficient answers are found. GES shows that the node vector size offers a good trade-off between search performance and bandwidth cost. Moreover, GES adopts automatic query expansion and local data clustering to improve search performance. Due to partial coverage problem nodes may left out form routing queries. No solution is proposed to prevent or reduce the effect of free riding.

*High Degree Random Walk Search (HHDRWS):* HHDRWS [54] is a search algorithm for improving the search efficiency and reducing unnecessary traffic in Gnutella. The scheme consists of two parts *(i) a resource information caching mechanism and (ii) a search technique*. The resource information caching mechanism is used for spreading resource information utilizing high degree nodes across the network. The search mechanism combines the random walk search and high degree walk search. The nodes that send query message choose neighbours according to random walk search method in the odd steps, and choose neighbours according to high degree walk search method in the even steps. Therefore, when a node started a query, it sent out 'b' query messages to an equal number of randomly chosen neighbours. After some nodes receiving the messages, they sent the messages to high degree neighbours according to high degree walk method in the second step. Then odd steps are random walk and even steps are high degree walk. HHDRWS achieves good success rates, reduces the search traffic, and balances the network. However, nodes are selected randomly for forwarding queries, hence the past performance of peers are not considered during searching. Moreover, the free riding and partial coverage problems are not addressed.

*Reinforcement Learning Based Search (ISRL):* ISRL [55] aims at locating the best path to desired files at low cost. It explores new paths by forwarding queries to randomly chosen neighbours. It also exploits the paths that have been discovered to reduce the cumulative query cost. Two models of ISRL are proposed: the basic ISRL for finding one desired file and MPISRL for finding multiple desired files. The ISRL learns the best path and adapts itself to the dynamic nature of nodes in the network. The random selection of nodes deteriorates the search performance. For choosing the target peers for routing queries, the behaviour of peers in the past is not considered.

*Distributed Search Technique (DST):* DST [56, 69] routes queries efficiently through well-performing nodes in the network for object discovery. DST employs a kind of reinforcement learning scheme known as Q-learning for choosing the target peers for routing queries. The learning process relies on a few number of attributes such as number of successful queries, number of results generated per query, hops visited, dynamic nature of nodes, Time-to-Live (TTL) etc of the peers. The search scheme achieves good load balancing, high hit ratio, low network traffic and adaptive behaviour. The most important features of DST are Q-learning based search, two-way load balancing, priority for specialized nodes, effective handling of duplicate messages, TTL enhancement and the utilization of past performance of nodes in routing queries. Peers in the network are classified as ordinary peers and power peers. Power peers and ordinary peers together join the search process. The indexes are maintained in data structures called Q-tables. A Q-value in a Q-table represents the behaviour of a node in the past. The Q-values in the Q-table are modified based on experience in routing. A two-way load-balancing strategy is employed for managing the query traffic. Routing queries simultaneously through ordinary nodes and power peers is a kind of load balancing being employed. This kind of load balancing helps to discover popular as well as unpopular objects from the nodes and balances the query load on various nodes. In the second type of load balancing strategy, using a mobile agent-based load balancing scheme, a highly loaded power peer redirects its future query load to least loaded power peer listed in the Q-table. DST effectively alleviates the partial coverage problem and the majority of nodes, excluding freeriders, are covered during searching. The duplicate messages are effectively routed to other nodes without generating more overhead. Specialized peers are given importance while routing queries since these nodes produce more number of results for queries on specialized areas. All the search operations terminate when either the result is found or the TTL is expired. However, the power peers can extend the search operation by means of TTL enhancement process and it is useful if the availability of objects are very low in the network. The performance of DST during simulation is compared with random walk and adaptive probabilistic search (APS) and DST shows improved success rate.

**Table 2. Advantages and Disadvantages of Informed Search Techniques**

| Search Technique | Query forwarding | Advantages | Disadvantages |
|---|---|---|---|
| Directed Breadth First Search (DBFS) | The query node sends the query message to a subset of its neighbours that will quickly return many high-quality results. These neighbours then forward the query message to all their neighbours just as in BFS. | i. Response time to a query is reduced as DBFS directs the queries towards peers that are more likely to respond with positive results.<br>ii. Improved success rate.<br>iii. Reduction in bandwidth and search cost a few peers are | i. Storage costs are increased, as each peer has to keep routing-relevant information about each of its neighbours.<br>ii. Fault-tolerance and reliability of the system is reduced, as there is no provision for data redundancy.<br>iii. DBFS is not flexible for easy |

| | | queried.<br>iv. Better scalability than BFS. | adaptation to peer departures. |
|---|---|---|---|
| Intelligent Search Mechanism | A peer estimates for each query, which of its peers are more likely to reply to the query, and propagates the query message to those peers only. | Shows better performance in environments, which exhibit strong degrees of query locality and where peers hold some specialized knowledge. | i. Search messages may be locked into a cycle and consequently fail to explore other parts of the network.<br>ii. Forwards the queries to same neighbour always , i.e. starvation for new peers |
| Adaptive Probabilistic Search | Searching is based on the deployment of k independent walkers and probabilistic forwarding. Each intermediate node forwards the query to one of its neighbours with probability given by its local index. Index values are updated using feedback from the walkers. | i. Peers eventually share, refine and adjust their search knowledge with time.<br>ii. Displays robustness when topology changes.<br>iii. Bandwidth-efficient<br>iv. Probabilistic selection of nodes instead of random selection. | i. Popular files could be located very fast, while other files could be hardly located.<br>ii. First discovered peer might be more used for future routing and might experience more load. At the same time, other peers, which are closer, could be ignored.<br>iii. Query load-balancing not addressed.<br>iv. Probabilistic selection of peers for routing and the node past behaviour in terms of various attributes such as bandwidth, storage, degree etc. are not considered.<br>v. All nodes are given equal status.<br>vi. Queries are routed through free riders.<br>vii. Duplicate messages are considered as failure states.<br>viii. Partial coverage problem exists. |
| Adaptive Resource based Probabilistic Search Algorithm | A node uses weighted probabilistic forwarding for query messages, varying the forwarding probability according to the popularity of the resource being searched and its own degree. Peers estimate the popularity of the resource in the network based on feedback from previous searches. | i. Lower search cost for popular files.<br>ii. Probabilistic selection of nodes, thus replaced random selection. | i. Forwarding probability is always static.<br>ii. Query load is not properly balanced.<br>iii. Exhibits partial coverage problem.<br>iv. Employs flooding for routing, thus creates message overhead.<br>v. Free riding problem is not addressed. |
| Activity Based Search | Routing of queries relies on the degree to which each neighbour has contributed to previous successful searches. | i. Better scalability<br>ii. A message per query is less than random walk. | The only parameter considered for decision is success ratio of previous searches. Other attributes such as storage, bandwidth, degree etc. are not utilised. |
| Preferential Walk | Employs a trust evaluation method to rate neighbours according to the feedback from previous searches. Neighbouring peers assign each other trust ranks. During routing, peers preferentially forward queries to the highly ranked neighbours. | i. Improved query success rate and response time.<br>ii. Minimises malicious behaviours by reducing the messages forwarded to low ranked neighbours. | i. Trust is probabilistically assigned, thereby accuracy may be reduced.<br>ii. Free riding problem and partial coverage issues are not addressed.<br>iii. Managing of duplicate messages not addressed. |
| Percolation Search | Using a probabilistic broadcast algorithm, a query message is relayed with probability just above the bond percolation threshold of the network. | i. Ability to answer complex queries using scalable resources.<br>ii. Overall traffic is reduced marginally. | i. Employs random walk for query implantation, and the peers available for a long time could be chosen more often, thus other nodes may be left out from searching, creating so called partial coverage problem.<br>ii. Solutions for handling free riders not provided. |
| Hybrid Periodical Flooding | Query mechanism is divided into several phases that are periodically repeated. | i. Based on multiple metrics the relay neighbours are selected.<br>ii. Alleviates partial coverage problem significantly. | Query processing is slow since searching relies on iterative deepening. |
| GIA | A search protocol based on biased a random walk that directs queries towards high-capacity nodes, which are typically best able to answer the queries. | i. The capacity constraints associated with each node is given priority while routing.<br>ii. Makes use of an active flow control scheme to avoid hotspots.<br>iii. Reduces the query-dropping rate in flooding.<br>iv. Query load is managed effectively. | i. Communication overhead.<br>ii. Low-capacity nodes may be omitted from routing and thus unpopular objects may not be queried. |
| Equation Based Adaptive Search | Utilises an estimate of the popularity of a resource to choose the parameters of random walk such that a targeted performance level is achieved by the search. | i. A popularity based search, which employs feedback obtained from previous searches.<br>ii. Significantly better than the non-adaptive random walk. | Unpopular objects are given low priority for querying. |
| Local Indices | Each peer maintains a local index of the content of all neighbouring peers within a hop distance called the radius of the index. | i. Shows good performance for topologies where only few nodes have very large numbers of neighbours.<br>ii. Reduced search and bandwidth costs and improved scalability. | i. To produce and uphold the indices at each node, extra steps must be taken every time a node joins or leaves the network<br>ii. Increased storage cost for maintaining index.<br>iii. Fault-tolerance and reliability are low. |
| Routing Indices | Each peer stores statistics of documents shared by it and documents shared by its neighbours and route query to a "good" peer. | i. Bandwidth efficient search<br>ii. Better performance than flooding-based search | i. Requires flooding, hence not always suitable for dynamic networks.<br>ii. Free riding and partial coverage issues not addressed. |
| Distributed Resource Location Protocol | Every peer only contacts peers in its neighbourhood as well as peers indicated in the Directory Cache. | The algorithm reduces the inefficiency of flooding by using a probabilistic message dissemination method combined with a caching mechanism. | i. The algorithm initially spends many messages to find the locations of an object.<br>ii. The algorithm backtracks if more results are needed. |
| Gnutella with Efficient Search | Organizes semantically relevant nodes into same semantic groups by using the notion of node vector. GES directs the query to the most relevant semantic groups. | Adopts automatic query expansion and local data clustering to improve search performance. | The issues such as free riding, partial coverage are not addressed. |
| High Degree Random Walk Search | Combines the high degree walk method and random walk method. Each node that wants to share its files spreads those files' information through high degree walk with limited steps. | i. Achieves good success rates.<br>ii. Reduction in search traffic. | i. TTL is always constant.<br>ii. Low degree nodes may be left out from searching.<br>iii. Past performance of nodes is not considered.<br>iv. Partial coverage problem exits.<br>v. Free riding problem not considered. |

| Reinforcement Learning Based Search | Each node is a distributed learner and the P2P network is its environment. Each learner iteratively estimates which next hop neighbour is the best one to forward a given query by trying them. | Learns the best path by reinforcement learning and adapts itself to system dynamics. | i. The random selection of nodes deteriorates the search performance.<br>ii. Load is not effectively balanced since queries are flooded. |
|---|---|---|---|
| Distributed Search Technique (DST) | The nodes in the network are classified as ordinary nodes and power nodes. K-walkers are propagated through the network. Queries are routed through nodes in the query cache. If the query keyword t is not found in the query cache, queries are routed through both ordinary nodes and power nodes based on past experience. | i. Applies Q-learning for selecting target nodes—thus avoids probabilistic and random selection of nodes for routing.<br>ii. Improved search performance.<br>iii. Reduction in message traffic with less hop distances.<br>iv. Adaptive behaviour.<br>v. Two-way load-balancing.<br>vi. Priority for specialized peers.<br>vii. Duplicate messages are managed significantly. | i. The messages from free riders are propagated through all the nodes.<br>ii. Required to maintain a few indices.<br>iii. The search performance relies on progress of learning. |

## 3. Replication Techniques

The objective of a replication technique is to improve availability and enhance system performance. P2P-based replication strategy is the topic of various active research projects and most of these projects emphasize replication in structured P2P systems. Only a few replication techniques are cited in the literature for unstructured P2P networks. This section briefly describes various replication schemes that have appeared in the literature. The replication techniques are generally classified (Fig. 5) according to site selection policy, replica distribution, erasure coded replication, and schemes for superpeer architecture. However, some techniques, which belong to one category, may possess the properties of another category. Eighteen replication techniques which belong to different categories are discussed in this paper. The Table 3 lists out the mode of replication, advantages and disadvantages of various replication strategies being employed in decentralised unstructured P2P networks.

The major features of replication algorithms for P2P systems are the criteria for the selection of suitable objects for replication, and selection of suitable sites for hosting the replica. If a node decides to replicate all the objects present in its shared directory to other nodes, it will increase the overhead in the network. The replica should be maintained in sites which are close to the source nodes to increase the search performance. The site selection policy of a replication technique decides where the replica should be stored. The number of sites may vary based on the replication scheme being employed. The second category of replication distributes replicas in the network based on number of copies to be distributed. In erasure code replication, a file is divided into b blocks. A variable amount of erasure code redundancy is then added to these blocks so that k > b blocks are obtained in total, with each block having the same size as before. The erasure-coded blocks are dependent on each other. Retrieving any b out of k blocks is enough to reassemble the original file. The erasure coded blocks are replicated to different sites based on replication rule. There are replication techniques exclusively for superpeer architecture such as ARRS and dynamic replication schemes.

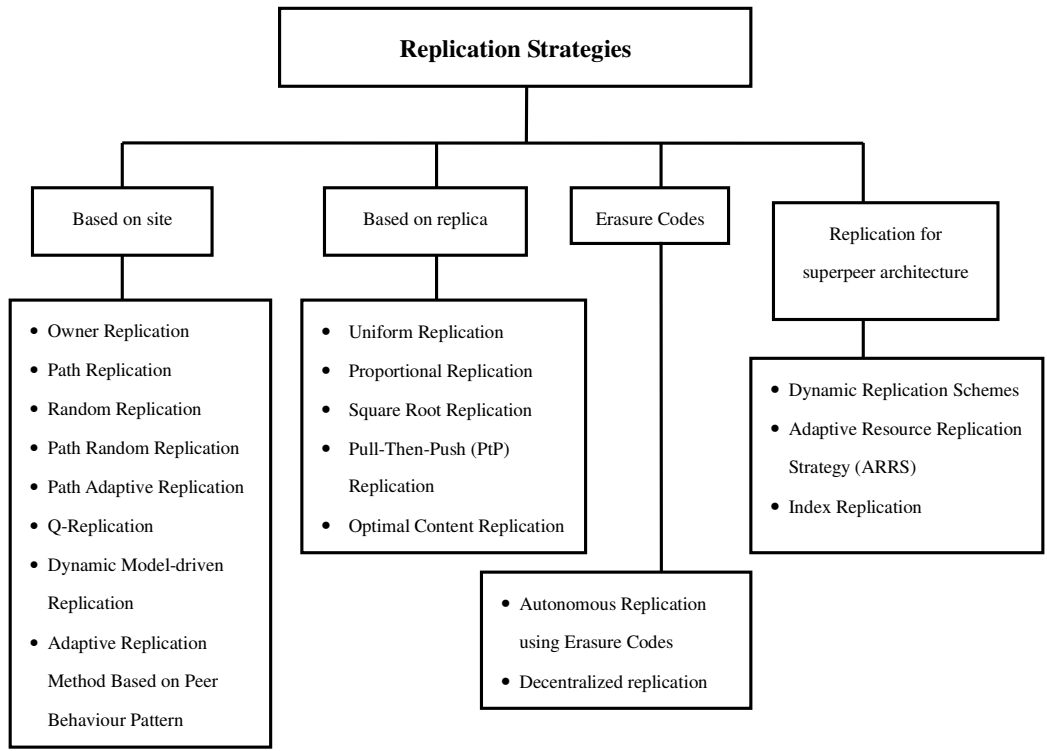

**Figure 5. Classification of replication techniques**

*Owner Replication, Path Replication and Random Replication:* Based on site selection policy the three schemes [8, 57, 58, 59] replicate the found object when a query is successful. The owner replication replicates an object only at the requesting node. The number of replicas increases in proportion to the number of requests for the service. Random replication distributes the replicas in a random order rather than following the topological order. If we use random forwarding n-walkers random walk, random replication is the most effective approach for achieving both smaller search delays and smaller deviations in searches. On the other hand, to carry out random replication, the peer must know the information of all the peers in the logical network. This is very difficult to implement since a peer only contains information about its neighbouring peers. The path replication creates copies of an object on all nodes on the path from the providing node to the requesting node and its implementation is less complex than the random replication. It has been shown in [8] that factor of improvement in path replication is close to 3 and in random replication the improvement factor is approximately 4.

*Path Random Replication and Path Adaptive Replication:* Path Replication [8] places replicas in all the peers on the path the requested data goes to the requesting peer. The number of replicas created can become very large, which eventually may be more than necessary to achieve the required search performance. Thus, some amount of the processing capability and storage capacity of the peers may be wasted, particularly on the few peers with a high degree. Path replication method coupled with a replication ratio is referred to as the path random replication method [60]. Each intermediate peer randomly determines whether or not the replica is created and placed there, based on the probability of the pre-determined replication ratio. The replication ratio is the ratio of the created replicas to all the intermediate peers on the path for each requested data. In path random replication, the requested data is replicated in each intermediate peer on the path with a specified probability, which is the same at any peer in the P2P network. This again increases load on high degree nodes that are frequently located in the data transmission path.

Path adaptive replication [60] is an alternative to path random replication that adaptively determines whether or not to create a replica depending on its storage capacity. Path adaptive replication determines the probability of the replication in each peer according to the predetermined replication ratio and its resource status. Path random replication out performs path adaptive replication in the average number of search hops. Path adaptive replication could further improve the load-balancing using each peer's local information on resource availability. The feedback information after replication is not collected and utilised for determining the target node for hosting the replica.

*Q-Replication:* This scheme is a modified form of replication scheme proposed in [61, 70]. Q-replication employs Q-learning for the autonomous replication of objects. It is autonomous because the decision to replicate an object to appropriate sites is taken autonomously by a node based on the past performance of peers in replicating objects. Thus in spite of constant changes to the connection, objects are highly available. Q-replication scheme maintains a Q-table which contains peer-IDs and corresponding Q-values of each peer. A Q-value represents how a peer has contributed to the replication activities in the past. As part of replication, a node receives a reinforcement signal from the target node which is intended for hosting the replica. The signal is translated into a reward. Parameters such as bandwidth, degree of the node, and storage cost are utilised. The Q-values are updated appropriately. The Q-replication process selects the target objects for replication based on their popularity. The popularity is computed in a unique way. After replicating an object, the Q-values corresponding to the nodes in the Q-table are updated using the various parameter values returned. A node which goes down frequently and maintains low values for bandwidth, degree, and available storage may produce small value for reward. These nodes show less performance in the replication process. Hence well-performing nodes receive high Q-values and the Q-values of nodes with low performance are reduced further. A shared object is replaced from a node to accommodate a new object by utilizing the popularity and the time at which the object was inserted into the shared directory of a node. The Q-replication, thus, distributes the popular objects to well-performing nodes in the network for improving the availability of objects and thereby contributes to the improvement of success rate and fault tolerance without depending on search paths.

*Dynamic Model-driven Replication:* This is a decentralized model, which is used for creating replicas dynamically in an unreliable P2P system [62]. Each peer in the system possesses a model of the P2P storage system that it can use to determine how many replicas of any file are needed to maintain desired availability. Each peer applies this model to information it has about system state and replication status of its files to determine if, when, and where new replicas should be created. Each node in the network is authorized to create replicas for the files it stores. The performance model compares the costs and the benefits of creating replicas of a particular file in certain locations. Although the technique is probability based, it is effective in predicting the required number of replicas in the system. The disadvantage of this approach is the replication process always relies on a resource discovery service to determine the number of replicas of the object exits in the P2P system. Replicas may get created unreasonably in the system if more than one node replicating the same file simultaneously.

*Adaptive Replication Method Based on Peer Behaviour Pattern:* Adaptive replication method based on peer behaviour pattern [63] uses the relevancy and usefulness of peers to determine how many replications should be made, and where to locate these replications. When a new document is registered at a peer, the peer replicates the document. The number of replications to be made depends on the relative usefulness of the neighbours of a given peer. The target peers are determined by a peer group and peer selection criterion, where a peer group is a candidate set of peers from which the target peers are selected. Four types of peer grouping are employed: placement on neighbours (PN), placement on inverted references (PIR), placement on relevant peers (PRP) and random placement. In PN method, the peer group is defined as the set of neighbours of the source peer. The neighbour peers are ranked in ascending order of their usefulness. In PIR method, a peer group is defined as the set of peers that access the source peer to obtain documents. PRP method defines a peer group as the set of peers whose queries are sent to the source peer. Random Placement defines the peer group as all peers joining the system. The target peer is chosen randomly from all peers. Using the PN method, query processing is effective when the network does not have relatively many peers. However, the performance degrades as the number of peers increase. The PN method experiences trouble with scalability. In the case of the random method, a favorite document is placed regardless of the distance between a query peer and the peer holding the document. However, the number of query results decreases as the number of peers increases. Neither the PIR method nor the PRP method is affected by an increasing number of peers, compared with the PN and random methods. For query results, the PRP technique shows superior performance than other techniques. Even though a node has several attributes, in this scheme the usefulness of a node is defined only using a single parameter. This makes the peer selection process for replication an imprecise one.

*Uniform, Proportional and Square Root Replication:* Optimal number of copies for each object in terms of the average search overhead per successful query are computed by these techniques [20, 64, 8]. In the uniform strategy, replications are uniformly distributed throughout the network. For each data object, approximately the same number of replicas are created. While this controls the overhead of replication, replicas may be found in places where peers do not access the files. In the proportional replication, the number of copies for each object is proportional to its query distribution. The higher the query rate of an object, the higher is the number of copies for that object. On the other hand, with proportional replication, although queries on popular data are processed efficiently, unpopular data search may take a long time, thus degrading the overall system performance. In the case of square root (SR) replication the number of replicas of a file 'i' is proportional to the square root of query distribution, qi. Optimal replication is attained when the number of replicas per item is proportional to the square root of their popularity. Uniform and Proportional strategies have been shown to have same search space.

*Pull-Then-Push (PtP) replication:* With PtP replication [58], after a successful search, the requesting node enters a replicate-push phase where it transmits copies of the item to its neighbours in order to obtain square root replication. Updating the replicas can be significantly improved through an update-push phase where the node that created the copies propagates any updates it has received using similar parameters as in replicate-push. The problem with SR replication is that it requires knowledge of the query rate for each item. To improve this, in PtP, after each successful search, the item is copied to a number of nodes equal to the number of probes. The creation of replicas is delegated to the inquiring node, not the providing node. The scheme consists of two phases. The pull phase refers to searching for a data item. After a successful search, the inquiring node enters a push phase, whereby it transmits the copies of the item to its neighbours in order to force creation of replicas. In order to reach SR replication, number of replicas equal to the number of probed nodes are created. The same algorithm is used for both the push and the pull phases, so that the push phase visits approximately the same nodes the pull phase visited. The TTL for push is equal to t-1, where 't' is the hop distance where the resource was found. Since replicas are placed on all the nodes probed, low performing nodes may also unnecessarily receive replicas.

*Optimal Content Replication:* Optimal content replication [65] is an adaptive, fully distributed technique that dynamically replicates content in a near-optimal manner. The optimal object replication includes a logarithmic assignment rule, which provides a closed form optimal solution to the continuous approximation of the problem. Two algorithms are proposed: a *Top-K LRU algorithm* and a *Top-K Most Frequently Requested (MFR) algorithm*. Top-K LRU algorithm replicates content on the fly without any a priori knowledge of object request patterns or node up probabilities. The idea behind the algorithm is as follows. Each object j has attractor nodes determined by the underlying P2P substrate. The object j tends to get replicated in its attractor nodes, which go up and down over time.

Queries for objects also tend to get sent to attractor nodes. Thus, a query for a particular object j tends to get directed to up nodes that likely have the object. Objects get replicated on the-fly when none of the top-K peers have the requested object. LRU lets unpopular objects remain in peers. When an unpopular object is requested, the object gets stored in one of the peers and remains there until it is ejected with LRU. If the object is very unpopular, it will likely not receive any requests during its halt in the peer, and hence waste storage space. The MFR algorithm is an alternative to Top-K LRU algorithm, which manages the storage effectively. MFR algorithm follows its own retrieval and replacement policy and makes high success rates that are very close to optimal.

*Autonomous Replication using Erasure Codes:* This scheme [66] uses randomized decisions extensively together with the application of erasure codes to tolerate autonomous peer actions. Each member of the P2P community hoards some subset of the shared files entirely on their local storage, called the member's hoard set, and pushes replicas of its hoard set to peers with excess storage using an erasure code. The basic steps of the algorithm are: (i) each member advertises the unique IDs of the files in its hoard set and the fragments in its replication store in the global index. Each member also advertises its average availability in the global directory, (ii) each member periodically estimates the availability of its hoarded files and the fragments in its replication store, (iii) periodically, say every Tr time units, each member randomly selects a file from its hoard set; the member does this by generating a random erasure coded fragment of the file and pushes it to a randomly chosen target, and (iv) the target accepts and saves the incoming fragment if there is sufficient free space in its replication store. If there is insufficient space, it either rejects the replication request or ejects enough fragments to accept the new fragment. Victims are chosen using a weighted random selection process, where more highly available fragments are more likely to be chosen. This method minimizes the bandwidth costs in accessing the files. The amount of replication of each file is proportional to the frequency of access to that file. The presence of a small number of highly available members can significantly reduce the replication necessary to achieve practical availability levels.

*Decentralized replication algorithms:* The decentralized replication algorithms [67] deal with storage allocation and replica placement. The process of storage allocation decides how many replicas can be produced for each file upon the limitation of storage space, and replica placement procedure decides the set of peers that are going to store those replicas of each file to achieve a reasonable level of file availability. To provide sufficient file availability, three heuristic algorithms-*a random algorithm, a group partition algorithm* that relies on peers' forming groups and highest available first (HAF) algorithm- *a greedy algorithm*, are proposed. HAF is based on an estimated system-level file availability target. The three replication schemes employ the erasure-coded blocks for replication. The random algorithm does not require any knowledge of peer availability, and gives each file the same stretch factor and equal opportunity in selecting peers. The group partition algorithm can achieve lower variance in file availability, hence may be a good choice if fairness of file availability is important. The greedy algorithm can achieve higher availability especially when peers share a small amount of storage space for replication and when high available peers in the system are rare. The success of the algorithms depends on the failure rate of peers in the network.

*Dynamic Replication Schemes:* Dynamic replication [68], which is used in superpeer P2P architecture, takes the cost of searching a data item and successfully replicates the most frequently accessed data files based on the access probabilities. Two novel techniques to share the load using replication techniques are proposed: periodic push-based replication (PPR) and on-demand replication (ODR). In PPR, the hosting superpeer periodically sends replicas of the most frequently accessed files to remote superpeers on the basis of global access frequency. A superpeer receiving a replica also informs its neighbouring superpeers about the replica through a restricted gossiping algorithm. ODR performs replication based on local access frequencies. A request for replication is initiated by a superpeer if the access frequency of a particular file reaches a predefined threshold. This technique allows superpeers to dynamically adapt to changes in access behaviour, however, it is greedy as each superpeer tries to perform replication based on its own needs rather than replicating from a global perspective as done in PPR.

*Adaptive Resource Replication Strategy (ARRS):* ARRS [71] is used for context-aware superpeer based P2P computing in a pervasive environment. The replication strategy utilises the resource request rate as the primary metric to start the replication process, and then adaptively replicate resources according to the properties of peers and the size of peer clusters. In addition, the strategy uses peer related information stored on superpeers to determine which peers should be selected to perform adaptive replications and where the resulting replicas should be stored. The ARRS reduces network delays while increasing resource success rate in comparison to commercial superpeer P2P systems and random replication strategies.

*Index Replication:* The objective of index replication technique [72] is to improve the search effectiveness for rare items, and reduce the bandwidth overhead incurred in superpeer based P2P networks. It explores the use of two-hop index replication to significantly improve the effective search space. In the two-hop index replication scheme, each node sends its index to all of its one-hop neighbours in its routing table. All of the one-hop neighbours, in turn, forward this index to all of their one-hop neighbours except the source node. This strategy effectively reduces to a two-hop flooding of indices around the nodes. Two variants of two hop index replication are used: SR replication and constant replication. In SR replication, each node performs one-hop replication. Superpeers then replicate the indices of their one hop neighbours to a random subset of their superpeer neighbours. Each super node's two-hop replica set size is equal to the square root of the number of its superpeer's neighbours. The advantage is reduction in the amount of replication and its cost. In constant replication, each superpeer does two-hop index replication to a constant number of superpeer neighbours. After each node does one hop index replication, superpeers propagate the index to only a constant number of their superpeers. This reduces indexing load on superpeers.

**Table 3. Advantages and Disadvantages of Replication Techniques**

| Replication Technique | Mode of replication | Advantages | Disadvantages |
|---|---|---|---|
| Uniform Replication | Replicates everything equally. | i. All resources are equally replicated regardless of their popularity.<br>ii. Replicas have utilisation rates proportional to their query rates. | i. Increases the search time.<br>ii. All objects have the same average search size. |
| Proportional Replication | The number of replicas is proportional to their popularity. The more peers demand a resource the more copies of the resource will be available. This is sometimes called Natural Replication. | i. Achieves good load-balancing with all replicas having the same utilisation rate.<br>ii. Average search sizes vary with more popular objects having smaller average search sizes than less popular ones. | i. Lesser popular items harder to find.<br>ii. The number of replicas created can become very large if path replication is employed. Thus replicas may be copied into low performing nodes on the path. |
| Square Root Replication | The number of replicas of a file is proportional to the square-root of query distribution. | i. Optimal replication.<br>ii. Performance lies in-between uniform and proportional replication techniques. | The search performance relies on the selection of suitable sites for hosting new replicas. Otherwise, replicas may be unnecessarily copied into infrequently queried nodes. |
| Owner Replication | The object is replicated only at the requester node once the file is found. | The number of replicas will increase in proportion to the number of requests for the service. | Takes a large amount of time to propagate replicas over the P2P network, thereby limiting the search performance for the requested data. |
| Random Replication | Stores the object randomly among nodes visited by the agents. | Creates the same number of replicas as of path replication. | i. The peer must know the information of all the peers in the logical network.<br>ii. Harder to implement because peers can only know information about its adjacent peers. |
| Path Replication | Stores the object along the path of a successful "walk". | Implementation less complex than random replication. | The number of replicas created can be very large. This large number of replicas occurs in high degree peers. Some amount of the processing and storage capacity of the peers may be wasted. |
| Pull-Then-Push replication | The scheme consists of two phases - pull phase and push phase. In addition, a procedure for replica update is discussed. The creation of replicas is delegated to the inquiring node. | i. Achieves square root replication.<br>ii. Replica update process achieves good replica placement and consistency with small message overhead. | The replicas are copied only to the neighbours of a node without considering their behaviour in the past. All the neighbours get a copy of the replica, which increases overhead, in the network. |
| Path random replication and Path | An extension of path replication. Path Random Replication, chooses | i. Allows each peer to determine whether or not to | i. Probability based.<br>ii. In path adaptive replication, |
| adaptive replication | the peers randomly with a predetermined replication ratio, and the another method.<br><br>Path Adaptive Replication makes a replica on a peer depending on how much storage is still available on it as well as the predetermined replication ratio. | create a replica based on its own local information, without any global information over the P2P network.<br>ii. Replicas are fairly distributed over all the peers, independently of their degree and replication ratio.<br>iii. Path Adaptive Replication achieves good search performance than Path Random Replication. | other node attributes such as bandwidth, node dynamics etc. are not considered for the selection of target nodes for hosting the replica.<br>iii. Queries in Path Adaptive Replication can more frequently discover the requested data in peers with a higher degree. |
| Optimal content replication | Based on logarithmic assignment rule | i. An adaptive replication technique.<br>ii. Dynamically replicates content in a near-optimal manner. | Past performance of nodes are not taken into account for selecting a suitable location for replication, which leads to resource wastage. |
| Adaptive replication method based on peer behaviour pattern | Uses the relevancy and usefulness of peers to determine how many replications should be made, and where to locate these replications. | i. When an original document is updated, the replications are updated.<br>ii. Query processing is effective.<br>iii. Performance does not decrease when more topics are used. | The usefulness of a node is defined only using a single parameter. This makes the peer selection process for replication an imprecise one. |
| Decentralized replication algorithms | Peers in this replication system adopt erasure code to replicate files. Three techniques are discussed. Each peer in this replication system is characterized by three parameters - online availability, a set of files that needs to be replicated, the amount of storage space that peer offers for replication purposes. | i. The *random algorithm* gives each file the same stretch factor and equal opportunity in selecting peers.<br>ii. The *group partition algorithm* achieves lower variance in file availability, hence may be a good choice if fairness of file availability is important.<br>iii. The *greedy algorithm* achieves higher availability especially when peers share a small amount of storage space for replication and when high available peers in the system are rare. | i. The issue of overwriting of same erasure coded block by other nodes in a peer is not addressed.<br>ii. Random selection of target peers; past performance of the peers is not considered.<br>iii. The method for accommodating new erasure blocks if storage exhausted is not provided. |
| Autonomous Replication using Erasure Codes | Adopt erasure code to replicate files. All replication decisions are made autonomously by individual members using only a small amount of loosely synchronized global state. | i. Minimises the bandwidth costs in accessing the files.<br>ii. The amount of replication of each file is proportional to the frequency of access to that file. | i. Random selection of target files without considering its popularity.<br>ii. A random erasure coded fragment of the file is pushed into a randomly chosen target. No other attributes of the target node is considered. |
| Q-Replication | Employs Q-learning for replicating objects.<br>Autonomous replication | i. Popularity based replication<br>ii. Replication relies on the past performance of nodes.<br>iii. Rank of a node is modified based on its contribution.<br>iv. Replication doesn't depend on a search mechanism. | Maintenance of indices creates overhead |
| Dynamic Replication Schemes | Employed for superpeer P2P architecture.<br><br>Replicates the most frequently accessed data files based on the access probabilities.<br><br>Two techniques proposed - Periodic Push-based Replication (PPR) and On-Demand Replication (ODR). | i. Periodic Push-based Replication (PPR) reduces the hop count to fetch the files.<br>ii. On-Demand Replication (ODR) provides adaptability to changes in access behaviour. | Suitable criteria for the selection of target nodes to host the replicas are not provided. |
| Dynamic Model-Driven Replication | Proposes a model to achieve a desired level of data availability, through which peers can decide when and where to replicate files. | i. Predicts the required number of replicas in the system.<br>ii. A node decides where to replicate a file using a performance model. | i. Files can be replicated only based on performance of a resource discovery mechanism.<br>ii. Extra replicas may get created in the event of more than one node replicating the same file simultaneously. |
| Index Replication | Used for Super peer P2P structure. Explores the use of multi-hop index replication.<br><br>Two variants of two-hop index replication are used: square-root replication and constant replication. | i. Improves the search effectiveness for rare items.<br>ii. Reduce the bandwidth overhead incurred in superpeer based P2P networks. | Only useful for replicating index in a super peer P2P architecture. |

## 4. Previous Work

A few survey papers on P2P systems and the information retrieval processes are mentioned in the literature. This section overviews previous survey papers related to resource discovery and replication in P2P networks.

In [74], authors present a survey and comparison of various structured and unstructured P2P networks. They categorized various schemes into these two groups in the design spectrum and discuss the application-level network performance of each group. Several examples of P2P applications in both the categories are briefed shortly. Finally, comparisons of two groups with different performance parameters are presented. The authors express their concerns regarding P2P overlay networks' virtual topology mapping to the physical network infrastructure. This has impact on the extra strain on the infrastructure which causes additional costs for the service providers. Another argument is that incentives implementation in P2P overlay services would also provide a certain level of self-regulatory auditing and accounting behavior for resource sharing. Further to that, the paper identifies, trust and reputation are vital for secured and reliable communications among the peers due to the intrinsic nature of peers. Moreover, proximity in P2P overlay routing is another significant factor in the routing decision for P2P overlay networks.

[64] presents a survey of replication algorithms for different distributed storage and content management systems ranging from distributed Database Management Systems, Service-oriented Data Grids, P2P Systems, and Storage Area Networks. Replication strategies in P2P Systems are classified into three - based on size of files (Granularity), based on replica distribution and the third category is based on replica creation strategy. Authors conclude that the main difference in replication protocols is due to consistency requirements. If an application requires rigorous consistency and has large numbers of update transactions, replication may diminish the performance as a result of synchronization requirements. However, if the application involves read-only queries, performance can be enlarged.

The two issues of P2P decentralized applications: discovery mechanisms, and trust management are surveyed in [74]. The survey identifies and defines key properties for each of these and also summarizes the efforts of the P2P community in addressing various attributes by categorizing and discussing relevant technologies and approaches.

## 5. Issues in P2P Search and Replication

From this theoretical survey on search and replication techniques for unstructured P2P networks, we have identified quite a few issues on searching and replication. The main challenge for information retrieval in P2P networks is to be able to guide the query to the sources that contain the most relevant answers in a fast and efficient way. The scalability issues and lack of centralized control pose difficult performance issues in unstructured P2P networks. A search technique should be able to decrease the number of messages sent per query while maintaining high success rate in short hop distances. The queries should be routed through less number of possible nodes to reduce overhead. The duplicate messages should be reduced and handled efficiently. Overloading of nodes causes reduction in their performance and so the search mechanism should provide good query load balancing among nodes. A large amount of peers in the P2P systems are freeriders and queries are seldom hit by those peers. Routing queries through free riders should be curtailed in order to decrease redundant messages. Not all the peers may be reachable during searching and this creates the so-called partial coverage problem. In order to alleviate this problem the queries should be routed appropriately. New neighbours, which are inserted into the neighbour list of a node, should be given preferences while routing queries. Nodes become frequently queried if they host popular objects. This leads to popularity based search and thus minimizes the messages and hops to be visited during search operations. Subsequently the load on those nodes gets increased. Some nodes in the network may host rare objects. Therefore, queries should be routed through those nodes to increase the success rate. Most of the search techniques employ probabilistic or rank based selection of nodes for routing queries in place of flooding. This demands an alternate mechanism for selecting nodes. Nodes, which are hosting objects on specialized areas, can produce more search results for particular queries and they should be given priority while routing queries. Another important issue in P2P network is nodes are connecting and disconnecting to the network at unpredictable times, hence the availability of resources is continually in flux. Therefore, a search scheme should adapt to rapidly changing environments. Peers are limited by their connection capabilities. This curtails the quantity of traffic the nodes can manage. The variety of search techniques discussed in the previous sections offer different solutions for various issues. Since each of these techniques do not present a complete solution independently, the research should focus on developing novel resource discovery schemes considering the above important issues.

Replicating objects to multiple sites has several issues such as selection of objects for replication, the granularity of replicas, and choosing appropriate site for hosting new replica [20]. The existing replication techniques address these issues differently. Excessive replication can cause wastage of network and peer resources and at the same time, scarcity of resources decreases the search success rate and increases the search delay. Two important aspects of replication—selection of file for replication and selection of site for hosting new replica—have a direct impact on the performance of the system. Suitable criteria should be followed for selecting a file for replication. If popular files are not replicated appropriately, overwhelming requests from peers can cause network congestions and slow download speed. Based on the location selection logic for hosting new replica, replicated copies should be placed in proximity to peers who are likely to request the resource. This allows peers to be able to search and find desired resources, and reduces delays taking place during search and downloading. The replication strategy should use different characteristics of peers such as available storage and their surrounding usage environment attributes such as network bandwidth to determine which peers should be selected to perform replications and where the resulting replicas should be stored. Majority of the existing replication methods only replicate objects to intermediate nodes between query node and target node. These replication schemes depend completely on the search path. Due to this, objects are unnecessarily replicated to low performing nodes on the search path. It is essential that the objects should not be replicated to low performing nodes since these nodes are not queried frequently by other nodes; excluding such nodes from replicating files can save bandwidth. In a network, many peers might have decided to replicate the same file at the same time. This should be managed; otherwise, the same file could be copied into nodes repeatedly. A replication scheme should be well designed to manage the frequent failure of nodes in the network to provide good success rate by maintaining replicas in other suitable peers. The various issues in replication demands more assertive replication approaches for unstructured P2P networks.

The growth of autonomic computing and bio-inspired approaches presents promising future for developing high-tech schemes for resource discovery and replication in unstructured P2P networks. Besides, techniques based on learning methods such as reinforcement learning, and Q-learning shall provide better results for information retrieval and availability improvement, especially for selecting suitable candidate peers for routing queries and hosting replicas. The development of semantic P2P networks, P2P multimedia streaming applications and mobile P2P systems increases the complexity of information retrieval process and these necessitate formulating new techniques based on the suitability of applications.

## 6. Conclusion

This article conducts a comprehensive theoretical survey of search and replication strategies in unstructured P2P networks. The classification of various techniques and a brief description of each of the techniques are discussed. The advantages and disadvantages of various search and replication techniques are summarized. Finally, various issues related to search and replication in unstructured P2P networks is presented.